%% file: main.tex
\documentclass[sigconf]{acmart}
\AtBeginDocument{%
  \providecommand\BibTeX{{%
    \normalfont B\kern-0.5em{\scshape i\kern-0.25em b}\kern-0.8em\TeX}}}

\setcopyright{acmcopyright}
\copyrightyear{2023}
\acmYear{2023}
\acmDOI{XXXXXXX.XXXXXXX}

\acmConference[SIGKDD '23]{29TH ACM SIGKDD CONFERENCE ON KNOWLEDGE DISCOVERY AND DATA MINING}{August 06--10, 2023}{Long Beach, CA}
\acmBooktitle{SIGKDD '23: ACM Special Interest Group on Knowledge Discovery and Data Mining, August 06-10, 2023, Long Beach, CA} 
\acmPrice{15.00}
\acmISBN{978-1-4503-XXXX-X/18/06}

\usepackage[ruled]{algorithm2e}
\usepackage{amsmath}
\usepackage{color}
\usepackage{balance}
\usepackage{ifthen}
\usepackage{multirow}
\usepackage{enumitem}
\usepackage[english]{babel} 

\usepackage[T1]{fontenc}
\usepackage[utf8]{inputenc}
\usepackage{babel}
\usepackage[font=small,labelfont=bf,tableposition=top]{caption}
\usepackage{booktabs}
\usepackage{threeparttable}
\usepackage{makecell}
\usepackage{amsfonts}
\usepackage{geometry}
\usepackage{subfigure}

\begin{document}

\title{Layout-aware Webpage Quality Assessment}


\author{Anfeng Cheng$^*$, Yiding Liu$^*$, Weibin Li$^{*}$, Qian Dong$^\dag$, Shuaiqiang Wang, Zhengjie Huang, Shikun Feng, Zhicong Cheng and Dawei Yin$^\mathsection$}
\affiliation{ \institution{Baidu Inc., Beijing, China \institution{$^\dag$Tsinghua University, Beijing, China}} \country{}}
% \affiliation{  \country{}}
% \affiliation{ \institution{$^\dag$ University of Chinese Academy of Sciences, Beijing, China} \country{}}
\email{{chenganfeng01, liweibin02, wangshuaiqiang, huangzhengjie, fengshikun01, chengzhicong01}@baidu.com}
\email{liuyiding.tanh@gmail.com, dq22@mails.tsinghua.edu.cn, yindawei@acm.org}
% \email{}
\thanks{$^*$ Co-first authors.}
\thanks{$^\mathsection$ Dawei Yin is the corresponding author}

\renewcommand{\shortauthors}{Cheng and Liu, et al.}
\newcommand{\etal}{\emph{et al. }}

\begin{abstract}
Identifying high-quality webpages is crucial for real-world search engines,
which can fulfil users' information need with less cognitive burden.
Early studies of \emph{webpage quality assessment} usually design hand-crafted features
that may only work on particular categories of webpages (e.g., shopping websites, medical websites).
They can hardly be applied to real-world search engines that serve trillions of webpages with various types and purposes.
In this paper, we propose a novel layout-aware webpage quality assessment model currently deployed in our search engine.
Intuitively, layout is a universal and critical dimension for the quality assessment of different categories of webpages. 
Based on this, we directly employ the meta-data that describes a webpage, i.e., Document Object Model (DOM) tree, as the input of our model. The DOM tree data unifies the representation of webpages with different categories and purposes and indicates the layout of webpages. To assess webpage quality from complex DOM tree data, we propose a graph neural network (GNN) based method that extracts rich layout-aware information that implies webpage quality in an end-to-end manner. 
Moreover, we improve the GNN method with an attentive readout function, external web categories and a category-aware sampling method. 
We conduct rigorous offline and online experiments to show that our proposed solution is effective in real search engines, improving the overall usability and user experience.
% The data and code are available at \url{https://github.com/NSF20/virt-gat}.
\end{abstract}


\maketitle

\input{sections/1_intro_new}
\input{sections/3_preliminaries}
\input{sections/4_method/method}
\input{sections/5_deployment}
\input{sections/6_evaluation/offline_evaluation}
\input{sections/6_evaluation/online_evalution}
\input{sections/7_case_study}
\input{sections/8_conclusion}

\bibliographystyle{ACM-Reference-Format}
\balance
\bibliography{main}

\end{document}

%% file: sections/1_intro_new.tex
\section{Introduction}
Search engines, such as Google and Baidu, plays an important role in fulfilling users' information need. Over the past decades, relevance modeling is the main concern of search engines, dedicated to putting the most relevant web content on top of the ranked results~\cite{zou2021pre,liu2021pre,shen2014learning,yin2016ranking}. However, the very fact that not all relevant contents are useful to users has become an increasingly serious symptom~\cite{mao2016does}, where relevant webpages with low quality would induce a significant cognitive burden on the user. For such kind of low-quality webpages, useful information is hard to be identified, and the users need to take extra effort to comprehend.
% understand the information to be conveyed. 

To reduce the cognitive burden, measuring the quality of webpages has become a critical concern, which can better benefit users with well-delivered information and improve the overall usability of a search engine. Intuitively, a webpage with a clear structure, tidy organization and concentrated delivery of crucial information is always preferable to one that only stacks content without proper presentation, even though each may contain similar information. For example, given webpages with comparable relevance, high-quality webpages should be ranked higher than its competitors.

Nevertheless, webpage quality assessment is a very challenging task in web search, due to the complexity and diversity of webpages in the era of web 2.0.
% A webpage with a clear structure, tidy organization and concentrated delivery of crucial information is always preferable to one that only stacks content without proper presentation, even though each may contain similar information. Therefore, an accurate assessment of webpage quality can facilitate a search engine to reduce the cognitive burden and more effectively provide useful information for users.
Previous attempts at webpage quality assessment mainly propose to manually design discriminative features~\cite{moustakis2004website,hasan2011assessing,caro2005data,gupta2021empirical}, where classification algorithms~\cite{gupta2021empirical,caro2007bayesian} are applied subsequently. 
However, modern search engines usually face trillions of webpages with various categories, where simple hand-crafted features and classification algorithms (e.g. Bayesian Networks~\cite{caro2007bayesian}) 
% may easily fail on such heterogeneous data, and 
can hardly capture the in-depth information that reveals the webpage quality. Moreover, most of them can only work on a particular category of webpages, e.g., shopping websites~\cite{cebi2013quality}, medical website~\cite{rafe2012qualitative}, web portals~\cite{caro2006defining}, and Wikipedia articles~\cite{hasan2009automatic}. 
They are infeasible to be effectively applied to real-world search engines that serve trillions of heterogeneous webpages.

To address the aforementioned limitations, we conduct the first work that investigates \emph{layout-aware} webpage quality assessment on real-world web data. 
% Given a certain piece of web content, the quality of a webpage is largely determined by its \emph{content layout}~\cite{moustakis2004website,caro2006defining}. 
The intuition is based on the findings that the quality of a webpage is largely determined by its \emph{content layout}~\cite{moustakis2004website,caro2006defining}, which is of a great influence on how users perceive textual and multi-modal content~\cite{wu2021lampret,zhang2020every,Xu2020LayoutLMPO,Xu2021LayoutLMv2MP,Huang2022LayoutLMv3PF}. Modeling in-depth layout information is appealing for webpage quality assessment in real-world web search scenarios. Yet, it is also challenging, where \textbf{two crucial research questions} need to be answered: 
\begin{itemize}[leftmargin=5mm]
    \item \textbf{RQ1:} \textit{How to capture the layout information of different categories of webpages in a unified manner?} 
    \item \textbf{RQ2:} \textit{How to encode webpage layout information for webpage quality assessment on large-scale heterogeneous webpages?}
\end{itemize}

% \textbf{RQ1:} \textit{How to capture the layout information of different categories of webpages in a unified manner?} \textbf{RQ2:} \textit{How to encode webpage layout information for webpage quality assessment on large-scale heterogeneous webpages?}

\begin{figure}[t]
	\centering
	\includegraphics[width=0.9\columnwidth]{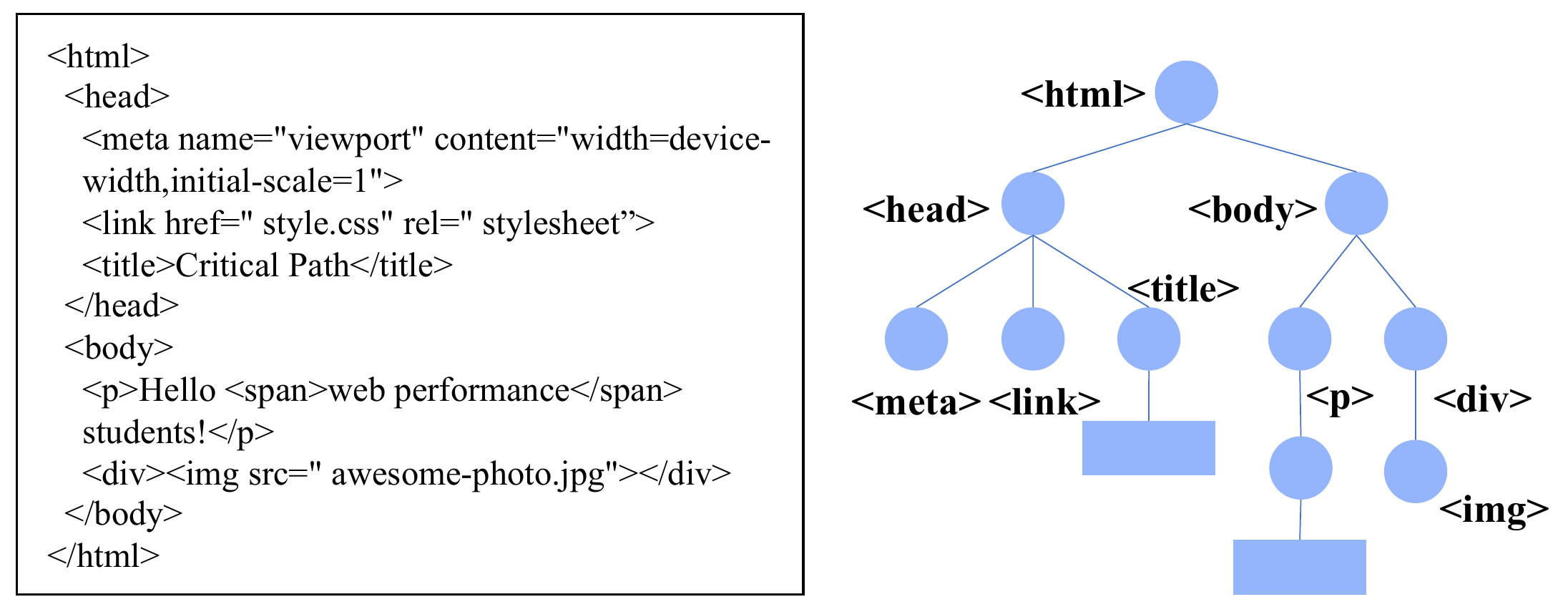}
	\caption{A toy example that shows the webpage layout represented by the DOM tree.}
	\label{fig:dom_tree}
\end{figure}

To answer \textbf{RQ1}, we propose to extract layout information of webpages from Document Object Model (DOM) data. Specifically, DOM is a cross-platform and language-independent interface that treats a webpage as a tree structure wherein each node is an object representing a content piece of the webpage\footnote{https://en.wikipedia.org/wiki/Document\_Object\_Model}. 
Figure \ref{fig:dom_tree} shows a toy example of the hierarchical structure of a DOM tree, which is converted from its HTML source code. Each node in the tree is an object that contains partial content of the webpage and is associated with different attributes that describe the content (e.g., type and size). Such data indicates rich hierarchical information on the content and its layout, and different categories of webpages can be represented in a unified manner. It is indisputable that inspecting the structure of DOM tree can help to measure the quality.

For \textbf{RQ2}, it is challenging as webpages in real search engines are highly diverse, where the modeling of layout information should be expressive to reveal the underlying patterns of heterogeneous DOM tree data. Recent advances in deep representation learning~\cite{bengio2013representation,lecun2015deep} have achieved great success on many web applications~\cite{shen2014learning,Chang2021WebQAMA,guo2021semantic}, and also sheds new light on our task at hand.
Notably, Graph Neural Networks~\cite{hamilton2017inductive,gilmer2017neural} has shown great performance in modeling structured text (e.g., word interactions)~\cite{zhang2020every,huang2019text,yao2019graph}, yet they are unexplored for complex DOM tree structure. 
Different from structure of textual document, the webpage layout represented by DOM tree is more complicated, which usually has hierarchical structure and the nodes usually have rich attributes. 
Existing methods that designed for text structure usually lack specialized consideration for the problem of quality assessment on DOM tree data, and thus might be unsatisfactory for real search engines.
% To solve \textbf{RQ2}
To this end, we propose the first GNN-based method to learn the underlying semantics of webpage layout in an end-to-end manner, based on which we further make several improvements to advance its performance on the task of webpage quality assessment.

To verify the effectiveness of our layout-aware webpage quality assessment model, we perform offline experiments on the dataset collected by the real-world search engine. Additionally, we deployed our model in the online ranking system and achieve good improvements.
Last but not least, the proposed solution is currently fully deployed in the online system of Baidu Search. 
To illustrate how layout-aware webpage quality assessment facilitates the overall usability of our search engine, we further present the details of the model deployment.
% , and we also present the details of how the solution is deployed.
% summary of contribution

Overall, our main contributions can be summarized as follows.
\begin{itemize}
    \item We develop the largest application of deep learning for the problem of webpage quality assessment, which significantly improves the overall usability of real-world search engines.
    \item We leverage DOM tree data and propose a GNN-based solution to learn the quality information of heterogeneous webpages in an end-to-end fashion.
    \item We present the deployment of webpage quality assessment model in the real production environment, which effectively and efficiently serves trillions of webpages with various categories and purposes.
    \item We conduct rigorous offline and online experiments before fully deploying the model online. The experimental results show that the proposed solution is effective to be applied in real-world search engines.
\end{itemize}

%% file: sections/3_preliminaries.tex
\begin{table}[ht]
    \caption{Commonly-used notations.}
    \centering
    \resizebox{0.48\textwidth}{!}{
    \begin{tabular}{l|l}
    \toprule \textbf{Notations} & \multicolumn{1}{c}{\textbf{Descriptions}} \\
    \midrule
    $\mathbf{G}_{p}=\{\mathbf{N},\mathbf{E}\}$ & A layout graph of webpage $p$\\
    \midrule
    $\mathbf{N}$ & The node set of graph $\mathbf{G}_{p}$ \\
    \midrule
    $\mathbf{E}$ & The edge set of graph $\mathbf{G}_{p}$ \\
    \midrule
    $\mathcal{F}=\{\mathbf{F}_t\}_{t=1}^{T}$ & The layout-related feature sets\\
    \midrule
    $T$ & The number of node type\\
    \midrule
    $\mathbf{F}_{t}=\{\mathbf{f}_{i}\}_{i=1}^{|\mathbf{F}_{t}|}$ & The feature set of node type $t$\\
    \midrule
    $\mathbf{f}_i$ & The layout-related feature\\
    \midrule
    $\mathbb{E}(\cdot)$ & The embedding of its input\\
    \midrule
    $\vec{h}_{n}^{(0)}$ & The initialized embedding of node $n$ \\
    \midrule
    $t_n$ & The node type of node $n$ \\
    \midrule
    $\boldsymbol{\beta}_p$ & The category of webpage $p$ \\
    \midrule
    $\vec{h}_{v}^{(0)}$ & The initialized embedding of virtual node $n$ \\\midrule
    $\sigma(\cdot)$ & An activation function\\\midrule
    $\alpha_{nm}$ & The attention score between nodes $m$ and $n$\\\midrule
    $e_{nj}$ & The attention coefficient between nodes $m$ and $j$\\\midrule
    $s_p$ & The predicted assessment score of webpage $p$\\\midrule
    $y_p$ & The manually assessment score of webpage $p$\\
    \bottomrule
    \end{tabular}}
    \label{tab:notation}
\end{table}

\begin{table*}[t]
    \caption{The considering aspects of rules and principles to score the layout of a webpage.}
    \centering
    \resizebox{1\textwidth}{!}{
    \begin{tabular}{l|l|l}
    \toprule 
    \textbf{Aspects} & 
    \multicolumn{1}{c|}{\textbf{Definition}} &
    \multicolumn{1}{c}{\textbf{Examples}} \\
    \midrule
    Interactive Experience            &
    Whether the webpage has interactive function &
    Click to call, swipe to browse pictures \\
    \midrule
    Paragraph            & 
    Ways to split the document into paragraphs &  Using different heading, special font color to layering \\
    \midrule
    Layout Design            &
    The overall design of the webpage's layout  &   
    many additional functions, various modules, font section size is appropriate \\
    \bottomrule
    \end{tabular}}
    \label{tab:rules}
\end{table*}

\section{Preliminaries}
In this section, we introduce the basic concepts and formalize the problem of webpage quality assessment. We summarize the commonly used notations in Table~\ref{tab:notation}.

\subsection{Layout-aware Webpage Quality}\label{sec: layout-aware_wq}
% According to the real search scenario, we formulate a brand new criteria to assess the quality of a webpage. 
Intuitively, high-quality webpages clearly provide useful information for users in common.
% The elements and key information of these webpages are always arranged in a reasonable manner. 
Specifically, given a set of webpages with comparable relevance under the same query, we consider the \emph{layout} (i.e. structure design, content presentation) as the \textbf{key dimension} of measuring webpage quality and improving user experience~\cite{moustakis2004website,caro2006defining}. Based on this, we can construct a set of rules and principles for annotating webpage quality and utilize human annotation as the objective of our proposed method. 

The considering aspects of rules and principles to score the layout of a webpage are shown in table \ref{tab:rules}, including interactive experience, paragraph and layout design.
We give the definition and some examples for each aspect in table \ref{tab:rules}. Webpages that meet the rules will get bonuses, and conversely, those who break the rules will be penalized.
Based on these principles, bonus and deduction rules can be formulated as webpages with reasonable \& beautiful layout or rich information will have an extra bonus, on the contrary, unreasonable \& chaotic layout or valueless information will be deducted.
Finally, annotators are required to score the given webpage 0 or 1 point based on the above rules and principles.
% The rules and principles for annotators are defined as the following:
% \begin{itemize}
%     \item 0 means poor layout. On the basis of ordinary pages, points will be deducted for various flaws.
%     \item 1 means ordinary layout. 1 point is common, and annotators are required to add or deduct on this basis.
%     \item 1.5 means better structure. a certain gain compared to ordinary layout.
%     \item 2 means gainful layout. The user experience of this layout is significantly better than most layouts.
% \end{itemize}

\subsection{Layout Graph}
% To represent quality information of webpages, 
To extract quality information from webpage layout, 
we construct a layout graph for each webpage based on its DOM tree. In particular, a layout graph is denoted as $\mathbf{G}_{p}=\{\mathbf{N},\mathbf{E}\}$ that contains a node set $\mathbf{N}$ and an edge set $\mathbf{E}$. Each node has a specific type (e.g., text, image and video), and is associated with several layout-related features $\mathbf{F}_{t}=\{\mathbf{f}_{i}\}_{i=1}^{|\mathbf{F}_{t}|}$. The features of different types of nodes are denoted as  $\mathcal{F}=\{\mathbf{F}_t\}_{t=1}^{T}$, where $T$ is the total number of node types. Besides, each layout graph is also associated with its webpage category, denoted as $\boldsymbol{\beta}_p$. The detailed construction process of layout graph is depicted in Section \ref{sec:graph_construction}.
% For the node level, several layout-related feature sets $\mathcal{F}=\{\mathbf{F}_t\}_{t=1}^{T}$ are carefully designed for each node type, such as text, image, video, and etc. $T$ is the total number of node type and each node type has a series of features, i.e., $\mathbf{F}_{t}=\{\mathbf{f}_{i}\}_{i=1}^{|\mathbf{F}_{t}|}$. Besides, each graph has a graph level feature category $\boldsymbol{\beta}_p$. 
% We slightly abuse the notation $\mathbb{E}(\cdot)$ to denote the representation of $\mathbf{f}_{i}$ and $\boldsymbol{\beta}_p$, i.e., $\mathbb{E}(\mathbf{f}_{i})$ and $\mathbb{E}(\boldsymbol{\beta}_p)$ indicate the representations of $\mathbf{f}_{i}$ and $\boldsymbol{\beta}_p$, respectively.

% \multicolumn{1}{p{4.5cm}}{}
\subsection{Webpage Quality Assessment}
Given a layout graph $\mathbf{G}_{p}=\{\mathbf{N},\mathbf{E}\}$, its features $\mathcal{F}=\{\mathbf{F}_t\}_{t=1}^{T}$ category $\boldsymbol{\beta}$ and $\mathbf{F}_{t}=\{\mathbf{f}_{i}\}_{i=1}^{|\mathbf{F}_{t}|}$, the task of webpage quality assessment is to estimate a score $s_p$ for a given webpage $p$ w.r.t. its quality, i.e,
\begin{equation}
    s_p = f_\theta(\mathbf{G}_{p}, \mathbf{F}_{t}, \boldsymbol{\beta}_p),
\end{equation}
where $f(\cdot)$ represents the quality model, and $\theta$ denotes its parameters. The scores should be consistent with users' perception of webpage quality, and reflect the rules and principles as we described above.
% In general, features in $\mathbf{F}$ can be embedded into Euclidean space, and the overall graph representation $\vec{h}_n$ is obtained from each feature representation $\mathbb{E}(\mathbf{f}_{i})$, combining with graph level feature $\mathbb{E}(\boldsymbol{\beta}_p)$ (i.e., category of webpage). Subsequently, graph representation is fed in the webpage quality assessment model to calculate the layout assessment score.

%% file: sections/4_method/method.tex
\section{Method}
In this section, we first present the overview of our model. Then, we describe the graph formulation process for a webpage, including the construction of the layout graph and feature pre-processing. After that, we present a GNN-based solution for webpage quality assessment.
% . To capture the fine-grained layout quality patterns from the constructed graph, several features are designed for nodes in this graph. 
% We leverage the layout graph and layout-related features
% , the node embedding could be initialized from the representation of these feature, where 
% and apply Graph Attention Network (GAT)~\cite{velivckovic2017graph} 
% is subsequently exploited 
% to propagate messages 
% to model the mutual influence 
% between adjacent nodes in the graph
% . We further design finalize a quality assessment score for each webpage.
% . The final assessment of a webpage is calculated through the aggregated graph representation, which is obtained by an expressive readout function using a virtual node. Lastly, we describe the details of optimization.

\subsection{Overview}
\begin{figure*}[t]
	\centering
	\includegraphics[width=15cm]{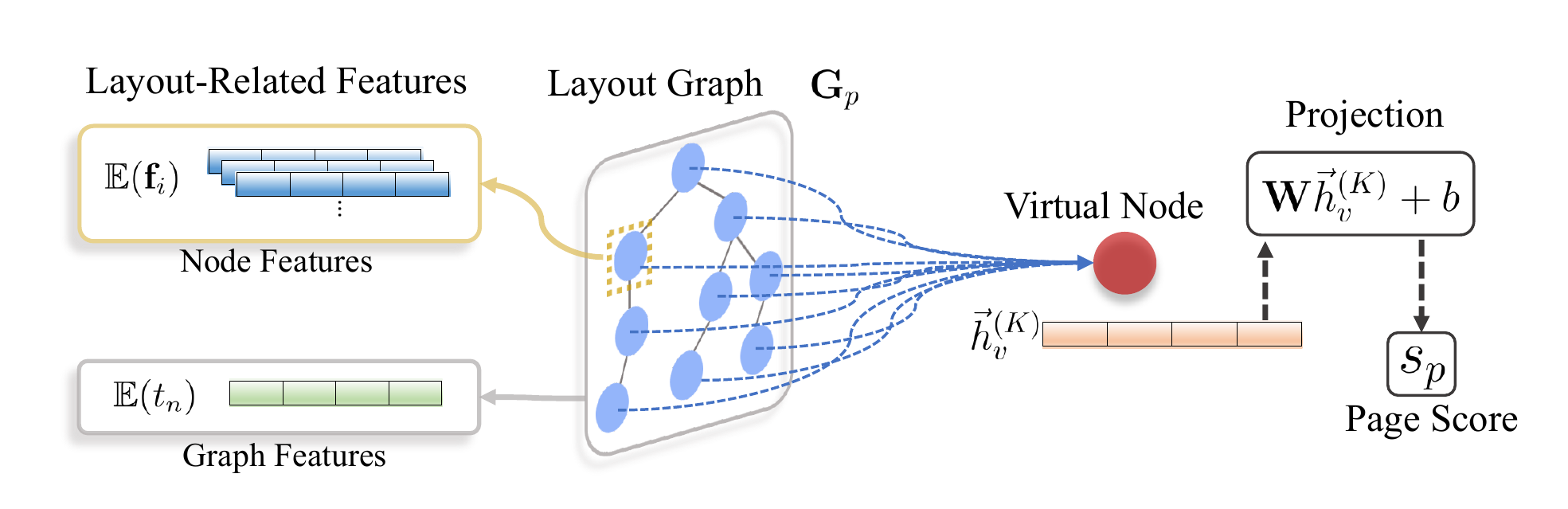}
	\caption{The illustration of message passing in our model. The red node represents the \textbf{virtual node} in the constructed layout graph, which is utilized for capture the graph-level information.}
	\label{fig:model}
\end{figure*}

Our solution mainly contains two components: \emph{layout graph formulation} (i.e., Section \ref{sec:graph_construction}) and \emph{quality assessment model} (i.e., Section \ref{sec:model}).
In layout graph formulation, we first leverage the layout information encoded in DOM tree to construct a layout graph $\mathbf{G}_{p}=\{\mathbf{N},\mathbf{E}\}$ for every webpage $p$.
% , where $\mathbf{N}$ and $\mathbf{E}$ represent nodes and edges respectively.
Then, two types of features are designed for the quality assessment, as depicted in Figure~\ref{fig:model}, including those associated with each node in the graph, as well as the category of the corresponding webpage (i.e., $\boldsymbol{\beta}_p$).

Next, we propose a quality assessment model that leverages Graph Attention Network (GAT) to perform expressive message passing between nodes in the layout graph. Both local and global structure information of the layout graph can be encoded in latent representations, which are exploited for the quality assessment task. Moreover, we improve the vanilla GAT model by 1) introducing an attentive readout function via the virtual node, 2) incorporating graph-level category information in the scoring function, and 3) alleviating the data imbalance problem that is common in real-world applications.

\begin{table}[t]
\caption{The summary of key features for each node type in our constructed graph.}
{
\begin{tabular}{l|l}
\toprule
\textbf{Classification} & \multicolumn{1}{c}{\textbf{Feature Name}} \\ 
\midrule
Location       & height, width, xpos, ypos, position type \\ 
\midrule
Content        & \begin{tabular}[c]{@{}l@{}}number of word, font size, font style,  \\ line height, font weight, alignment\end{tabular} \\ 
% delete direction
\midrule
Layout         & \begin{tabular}[c]{@{}l@{}}border, padding, margin, visibility, \\ display style, outline style, outline width\end{tabular}  \\ \midrule
Others          & tag name, webpage category  \\ 
\bottomrule
\end{tabular}}
\label{table:features}
\end{table}

\subsection{Layout Graph Formulation} \label{sec:graph_construction}

\paragraph{\textbf{Graph construction}}
% The layout information underlie in the DOM tree has a significant effect on the quality assessment of webpages. 
The content layout has been viewed as one of the most critical dimensions for measuring webpage quality~\cite{moustakis2004website}. To formulate layout information for various categories of webpages, we first construct layout graph based on DOM tree.
% Therefore, it is a core of our assessment system that how to extract the layout graph from the DOM tree. 
In particular, we leverage HTML parser Beautiful Soup \footnotemark[1]\footnotetext[1]{We parse webpages with the python library: \url{https://www.crummy.com/software/BeautifulSoup/bs4/doc/}} to parse the source code of a webpage, identifying the hierarchical structure of the webpage. Then, Depth First Search (DFS) is used for exacting adjacency relationships from the DOM tree.
Specifically, we recursively record the nodes and the corresponding edges between parent and child nodes in the DOM tree, as shown in Algorithm~\ref{alg:construction}. The layout graph $\mathbf{G}_p$ of webpage $p$ can be expressed by the exacted nodes $\mathbf{N}$ and their relations $\mathbf{E}$.

\paragraph{\textbf{Virtual node}} It is worth noting that, we also include a global \emph{virtual node} that connects to all the other nodes in the graph (as shown in Figure \ref{fig:model}). It can be viewed as a super-hub \cite{Ying2021DoTR} of the layout graph, which could be useful to aggregate the global information, and serves as hyperlinks that connect any two nodes in the layout graph. As such, we can capture global information of the given graph via the virtual node.
% The whole process of graph construction is illustrated in Figure~\ref{fig:graphconstruction}.
\input{sections/4_method/alg_graph_construction}

\paragraph{\textbf{Feature pre-processing}}
% delete reading direction
To capture the layout information of the webpage, we design a series of features for each node type. Taking the text node as an example, font style, font size, alignment and position in webpage are all represented by learnable embedding. The detailed list of features is presented in Table~\ref{table:features}.

More specifically, for continuous features (e.g., height, line height and margin), a non-uniform interval division strategy is employed to divide the continuous interval into several buckets, which can ensure that there are enough training samples in a single bucket. The uniform division of the whole interval leads to the data sparse issue since the continuous features typically obey a long-tail distribution.
Discrete features (e.g., font style, display style and tag name), are falling on a divided interval are mapped into a corresponding bucket, and this bucket is assigned a learnable embedding to represent the characteristics of its interval. 
% For discrete features, we naturally assign an embedding for a class. 

In addition to the node-level features, graph-level feature embedding is introduced to the layout graph (i.e., webpage category) to provide the model with the ability to perceive different categories of webpages, which is vital to the quality assessment. 
\textbf{One reason} is that the same webpage category has a similar structure. With the development of webpage makers (like Dreamweaver, and Google Web Designer),  large amounts of webpages are generated from templates and almost in the same layout. Therefore, with this graph-level embedding, the predicted assessment score shall be more robust in the online search engine.
\textbf{Another reason} lies in that different webpage categories have different criteria for quality assessment. For example, a succinct and well-organized document layout without distracting pictures is preferred on a search page, but for a portal, a document layout with pictures and text is considered to be better. 
In summary, it is meaningful and important to take the graph-level feature embedding into account for the layout graph.

\subsection{Quality Assessment Model} \label{sec:model}
% After layout graph construction and initialization, how to properly
Given the constructed layout graph associated with rich features, the key to webpage quality assessment is to expressively reveal salient patterns underlying the graph. In particular, we consider two types of relationships in the graph that could be discriminated for the task: \textbf{Local relationships}. Intuitively, the relationships between adjacent nodes in the layout graph are important to reveal content quality. For example, a node with \texttt{<image>} tag is usually the illustration of its adjacent (e.g., parent) node with \texttt{<div>} tag, which contains textual description. The interaction of the two nodes indicates the web content has both visual and textual presentation, forming a strong signal of high-quality content. \textbf{Global relationships}. Another important insight is that the relationships between local content and global layout should also be considered. For example, a node with textual description might be critical in a news article but is less important in a video webpage, whose quality largely depends on the node that contains the video.
% \begin{itemize}
%     \item 
%     \item 
% \end{itemize}

% should have a more significant impact on the layout quality than other adjacent nodes. 
% An image with a caption containing appropriately words is typically appeared in a well-structured page. Such layout quality pattern can be carefully considered through adequate interaction between nodes.

\paragraph{\textbf{Attentive message passing}}
To achieve this, we leverage graph neural networks that are promising to capture such complicated patterns.
% propagate the message in this graph becomes a crucial issue, since the layout quality patterns need to be sufficiently captured from the interactions (i.e. message passing) between nodes. 
% Therefore, the message passing module should have the capability of discriminating adjacent nodes.
In particular, we utilize the Graph Attention Network (GAT)~\cite{velivckovic2017graph} to model the interactions between nodes in the layout graph, where the modeling of node relationships can be viewed as message passing~\cite{gilmer2017neural} among nodes.
% , as the messages from adjacent nodes could be discriminated through the attention mechanism. 

In particular, the architecture of GAT is composed by stacking multiple graph attention layers, each of which can be defined as
\begin{equation}
\label{eq:GAL}
\vec{h}_{n}^{(k+1)}=\sigma\left(\sum_{m \in \mathcal{N}_{n}} \alpha_{n m} \mathbf{W}_{1}^{(k)} \vec{h}_{m}^{(k)}\right),
\end{equation}
where $\sigma(\cdot)$ is an activation function and $\alpha_{n m}$ is the attention value between node $n$ and node $m$. Here, $\vec{h}_{n}^{(k)}$ represents the embedding of node $n$ in the $k$-th layer.
% , and $\vec{h}_{n}^{(0)}=\vec{h}_{n}$. 
%  as defined in Eq.(\ref{eq:featurePooling})
The attention value $\alpha_{n m}$ is learned to selectively propagate information from neighbour node $m$ to node $n$, and a node can attentively interact more with its important neighbours than those trivial ones. Formally, the attention value can be defined as
\begin{equation}
\label{eq:attention}
\alpha_{n m}=\operatorname{softmax}_{m}\left(e_{n m}\right)=\frac{\exp \left(e_{n m}\right)}{\sum_{j \in \mathcal{N}_{n}} \exp \left(e_{n j}\right)},
\end{equation}
where the logits $e_{n j}$ is computed as 
\begin{equation}
e_{n j}=\sigma\left(\mathbf{W}_{3}^{(k)}[\mathbf{W}_{2}^{(k)} \vec{h}_{n}^{(k)}\|\mathbf{W}_{2}^{(k)} \vec{h}_{j}^{(k)}]\right).
\end{equation}
Here, we use $\|$ to represent the concatenation operation, and $\mathbf{W}_{2}^{(k)}$ and $\mathbf{W}_{3}^{(k)}$ are the weight matrices of the linear transformations at the $k$-th layer. Note that the weight matrices are shared across different nodes in a single graph attention layer.
% Following the vanilla GAT~\cite{velivckovic2017graph}, 
% shared linear transformations, parametrized by weight matrices $\mathbf{W}_{2}^{(k)}$ and $\mathbf{W}_{3}^{(k)}$, are applied to every node to obtain sufficient expressive power and the $attention$ $coefficient$, respectively. 

After $K$ times of message passing, the layout-aware patterns could be captured by node interactions (as defined in Eq. (\ref{eq:GAL})) within $K$-hops. It is worth noting that the virtual node also plays an important role during the message passing process. The virtual node offers a pathway for nodes’ interaction with considering the global interactions in the graph, which is critical for the quality assessment task. Overall, the GAT-based message passing framework is able to comprehensively model both local and global relationships for the final task.
% Note that the virtual node is connected to all the nodes in the graph, and thus both local and global relationships between nodes  and 
% To investigate the performance gain brought by virtual node, we remove $\vec{h}_{v}$ and its corresponding edges from $\textbf{G}_p$ and designed a traditional readout out function by re-write Eq.(\ref{eq:attnReadout}) as

\paragraph{\textbf{Readout function}} 
To compute the final quality score, we define the readout function as mean-pooling~\cite{hamilton2017inductive,wu2020comprehensive} to summarize all node representations as the final graph representation, and subsequently adopt a linear layer as
\begin{equation}
\label{eq:traditionalReadout}
    s_p = \mathbf{W}\operatorname{mean\_pooling}(\vec{H}_{\mathbf{N}}^{(K)})+b,
\end{equation}
where $\vec{H}_{\mathbf{N}}^{(K)}$ is the set of node representations in $K$-th layer of GAT. Alternatively, we can apply a more reasonable readout function, which is to use the representation of the virtual node as the final graph representation, and rewrite Eq. (\ref{eq:traditionalReadout}) as
\begin{equation}
\label{eq:attnReadout}
    s_p = \mathbf{W}\vec{h}_{v}^{(K)}+b,
\end{equation}
where $\vec{h}_v^{(K)}$ is the virtual node representation in $K$-th layer (i.e. the last layer) of the model. 
% Through this virtual node, any two nodes can interact with each other within two hops. 
In such case, the aggregation on the virtual node can be viewed as an attentive readout function, which has the capability of distinguishing the impact of different nodes in the graph for the final task.

\paragraph{\textbf{Category-aware quality assessment}}
% importance
% Q1: Why the category information of webpage plays key role in webpage quality assessment? I wanna some statistics, papers or theory.
% Q2: 上面是说Category信息好，下面说Category信息有bias，需要尽可能避免，我觉得应该别删。
% \xyq{Need Discussion.}
% As presented in Section \ref{sec: layout-aware_wq}, the category information of webpage plays a key role in webpage quality assessment.
The quality score defined in Eq. (\ref{eq:attnReadout}) is based on rich information aggregated from nodes. However, graph-level information is critical yet not incorporated. Therefore, we further improve Eq. (\ref{eq:attnReadout}) with the category information of webpage. In particular, we denote the category embedding of a given webpage $p$ as $\mathbb{E}(\boldsymbol{\boldsymbol{\beta}}_p)$, and further rewrite Eq. (\ref{eq:attnReadout}) as 
\begin{equation}
\label{eq:cateAttnReadout}
    s_p = \mathbf{W}(\vec{h}_{v}^{(K)} + \mathbb{E}(\boldsymbol{\boldsymbol{\beta}}_p))+b.
\end{equation}
% \begin{equation}
% \label{eq:featurePooling}
%     \vec{h}_{n} = \frac{\sum_{t=1}^{T}\sum_{i=1}^{|\mathbf{F}_{t}|} \mathbb{E}(\mathbf{f}^{n}_{i})}{\sum_{t=1}^{T}\sum_{i=1}^{|\mathbf{F}_{t}|} \mathrm{1}}+\mathbb{E}(\boldsymbol{\boldsymbol{\beta}}_p),
% \end{equation}
Note that the category embedding $\mathbb{E}(\boldsymbol{\boldsymbol{\beta}}_p)$ has the same dimensionality as the graph embedding $\vec{h}_{v}^{(K)}$, such that the embeddings could be summed for the final assessment. 
% where $\mathbb{E}(\cdot)$ means the embedding of its input. For node $n$ belonging to type $t_n$, other types of feature (i.e. $t \neq t_n$) are represented by a padding value. Consequently, the virtual node $\vec{h}_{v}$ is initialized from the addition of mean-pooled padding values and webpage category embedding $\mathbb{E}(\boldsymbol{\boldsymbol{\beta}}_p)$.

\begin{figure*}
    \centering
    \includegraphics[width=16cm]{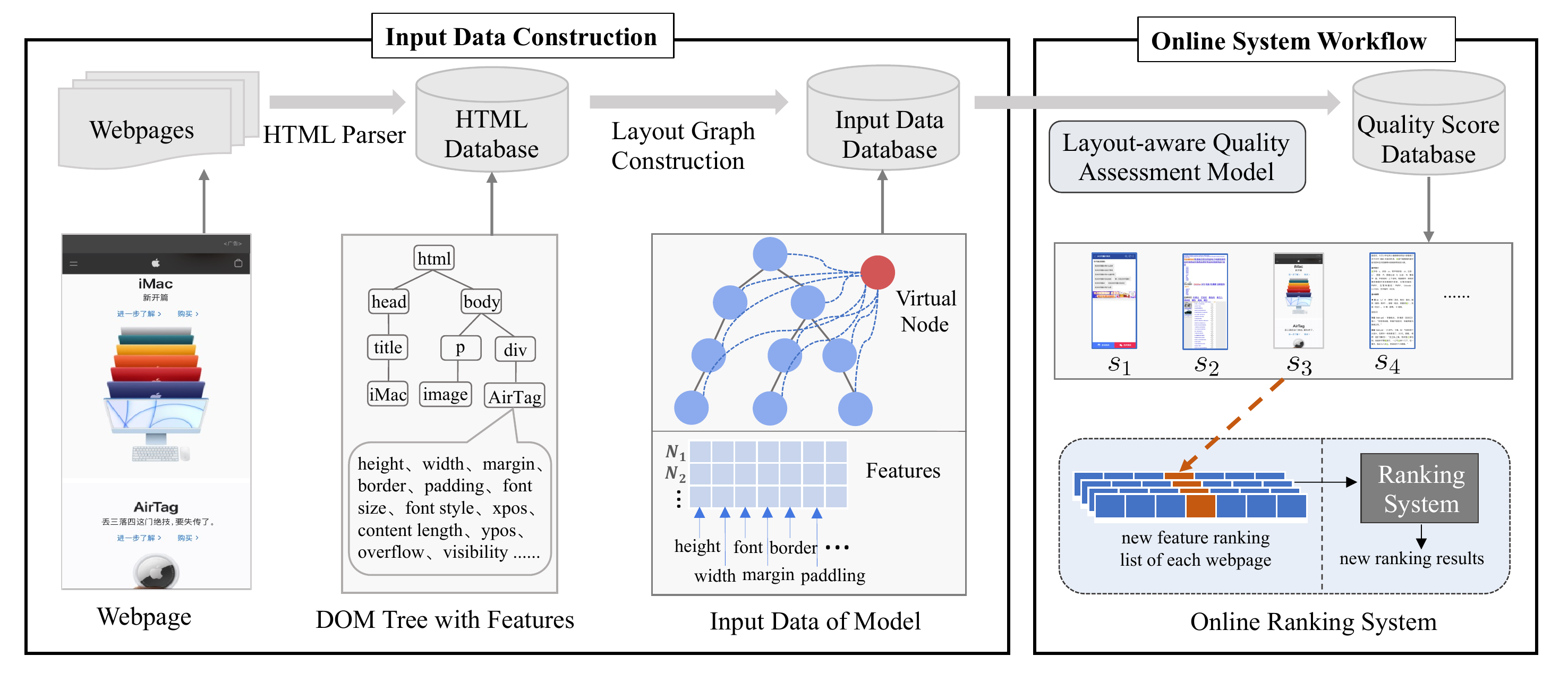}
    \caption{The overview of deployment in online ranking system.}
    \label{fig:deployment}
\end{figure*}

% \subsection{Optimization}
% \subsubsection{Assessment Criteria}
% From interactive experience, richness, content paragraphs and structure design, we formulate a brand new criteria to assess the layout of a webpage, and score the layout into four grades: 0, 1, 1.5 and 2. Based on this criteria, the largest manual annotation dataset of webpage layout is constructed. The score does not depend on the quality of the content in the webpage, such as web spam, inappropriate answers, etc. Bonus and deduction items are described as follow:
% \begin{itemize}
%     \item For the interaction design, a high-quality experience significantly different from ordinary webpages is a bonus, while interactive elements that cannot be used normally, or have poor interactive experience and even lack necessary interactive elements are deducted.
%     \item For the main body of webpage, a rich content containing many elements is bonus, while lack or empty of content is deducted.
%     \item For the content paragraphs, title and body are distinguished by size, font and color are bonus, while points are deducted if there is no discrimination in content paragraphs.
%     \item For the overall layout of the webpage, containing additional functions and plates with reasonable and beautiful layout are bonus, on the contrary, points are deducted.
% \end{itemize}
% The score of ordinary webpages is 1. A bonus or deduction item adds or subtracts 0.5 point respectively. All deduction items take precedence over bonus items. The assessment score is set to 0 if there is a deduction item and the maximum score is 2.

\paragraph{\textbf{Category-aware data sampling}}
As the graph-level category embedding is introduced in Eq.(\ref{eq:cateAttnReadout}) to perceive different categories of webpages, the bias in different categories may affect the prediction of models. In particular, some webpages are highly similar in layout, such as some popular question-answering websites, which are generated from templates. Such webpages typically have similar layout scores. Consequently, the predicted assessment score may be dominated by the category-aware embedding (i.e. graph level embedding). To alleviate this issue, a category-aware sampling strategy is employed.
% Specifically, we group the last two grades 1.5 and 2 into one class and 0 into another class. 
Up-sampling is utilized to balance the number of two classes, based on which the bias could be mitigated and our model 
could learn a distinguishable quality assessment score for a single category of webpages.

\paragraph{\textbf{Optimization objective}}
After up-sampling, the model could be optimized through Mean Squared Error (MSE) loss. It can be defined as
\begin{equation}
J=\frac{1}{P} \sum_{p=1}^{P}\left(y_p-s_p\right)^{2},
\end{equation}
where $P$ is the total number of training samples after up-sampling and $y_p$ is the annotated layout score of webpage $p$.

%% file: sections/4_method/alg_graph_construction.tex
\begin{algorithm}[!tb]
\caption{Layout Graph Construction}
% Algorithm}
\label{alg:construction}
\KwIn{HTML DOM tree $\mathbf{R}_p$ of webpage $p$}
\KwOut{layout graph $\mathbf{G}_p=\{\mathbf{N},\mathbf{E}\}$ of webpage $p$}
\hrulefill 

\% recursive graph construction;\\
% \FuncSty{${\mathrm{GraphConstruction1}}(\mathbf{root})$} \Begin{
%     $\mathbf{N}_r=\{\mathbf{root}.node\}$;\\
%     $\mathbf{E}_r=\{\}$;\\
%     \For{$\mathbf{child} \in \mathbf{root}.children$}{
%         $\mathbf{N}_c = \mathbf{child}.node$;\\
%         $\mathbf{N}_r=\mathbf{N}_r \cup \mathbf{N}_c$;\\
%         $\mathbf{E}_c = (\mathbf{child}.node, \mathbf{root}.node)$;\\
%         $\mathbf{E}_r=\mathbf{E}_r \cup \mathbf{E}_c$;\\
%         $\mathbf{N}_c, \mathbf{E}_c = \mathrm{GraphConstruction1}(\mathbf{child})$;\\
%     }
%     return $\mathbf{N}_r,\mathbf{E}_r$
% }

\FuncSty{${\mathrm{GraphConstruction}}(\mathbf{root})$} \Begin{
    $\mathbf{N}_r=\{virtual\_node, \mathbf{root}.node\}$;\\
    $\mathbf{E}_r=\{(virtual\_node, \mathbf{root}.node)\}$;\\
    \For{$\mathbf{child} \in \mathbf{root}.children$}{
        $\mathbf{N}_c = \mathbf{child}.node$;\\
        $\mathbf{N}_r=\mathbf{N}_r \cup \mathbf{N}_c$;\\
        $\mathbf{E}_c = \{(virtual\_node, \mathbf{child}.node),$\\ 
        $(\mathbf{child}.node, \mathbf{root}.node)\}$;\\
        $\mathbf{E}_r=\mathbf{E}_r \cup \mathbf{E}_c$;\\
        $\mathbf{N}_c, \mathbf{E}_c = \mathrm{GraphConstruction}(\mathbf{child})$;\\
    }
    return $\mathbf{N}_r,\mathbf{E}_r$
}
$\mathbf{G}_p = \{\mathrm{GraphConstruction(\mathbf{R}_p)}\}$;\\

\end{algorithm}

% \% recursive graph construction;\\
% \FuncSty{${\mathrm{GraphConstruction}}(\mathbf{root})$} \Begin{
%     $\mathbf{N}_r=\{\mathbf{root}.node\}$;\\
%     $\mathbf{E}_r=\{\}$;\\
%     \For{$\mathbf{child} \in \mathbf{root}.children$}{
%         $\mathbf{N}_c, \mathbf{E}_c = \mathrm{GraphConstruction}(\mathbf{child})$;\\
%         $\mathbf{N}_c = \mathbf{child}.node$;\\
%         $\mathbf{E}_c = (\mathbf{child}.node, \mathbf{root}.node)$;\\
%         $\mathbf{N}_r=\mathbf{N}_r \cup \mathbf{N}_c$;\\
%         $\mathbf{E}_r=\mathbf{E}_r \cup \{(\mathbf{child}.node, \mathbf{root}.node)\} \cup \mathbf{E}_c$;
%     }
%     return $\mathbf{N}_r,\mathbf{E}_r$;
% }

%% file: sections/5_deployment.tex
% \begin{figure*}
% 	\centering
% 	\includegraphics[width=18cm]{pic/deployment.pdf}
% 	\caption{The illustration of input data construction process.}
% 	\label{fig:deployment}
% \end{figure*}
\section{Deployment}\label{sec::deployment}
In this section, we show how the layout-aware webpage quality assessment model be applied to our online ranking system. We first introduce the input data construction process of the quality assessment model and then present the general picture of the quality score working in the ranking system. The overview of deployment is shown in Figure~\ref{fig:deployment}.

\subsection{Offline Input Data Construction}
% \noindent\textbf{Offline input data construction.}
In the left component of Figure~\ref{fig:deployment}, we present the process of input data construction for our model. Firstly, each webpage on the world wide web will be parsed through our HTML parser. All features of the HTML are stored in a database. Secondly, we construct the layout graph based on DOM tree and extract the features needed for quality assessment model using the algorithm defined in Algorithm~\ref{alg:construction}. Note that this process runs offline, it can significantly reduce the computing time of the online search system.

We also list the features which are used in our webpage quality assessment model, details are shown in Table~\ref{table:features}. We classify the features into three main categories w.r.t., location, content, and layout according to the different roles they play in building webpage. Category location is the primarily feature that locates the position of elements in the webpage e.g., height, width and position type. Category content contains text-related features e.g., the number of words, font style, and line height. Category layout is a feature that controls the layout of elements, e.g., border, padding, and margin. In addition, we add tag name, natural categorical information, and webpage category, which is used to balance the distribution of train data under different webpage forms.

\subsection{Online System Workflow}
% \noindent\textbf{Online system workflow.}
The online system workflow is presented in the right component of Figure~\ref{fig:deployment}. Our ranking system contains a wide variety of webpage features, where quality is one of the most important factors. To apply our layout-aware webpage quality assessment model in our online retrieval system, the new quality scores need to be loaded into the retrieval feature list. The online ranking system only needs to load the new quality assessment score and apply it to obtain the new ranking results with respect to the new ranking webpage list, which is shown in the lower left area of the online component. Note that, the quality assessment scores of all webpages are calculated offline and are independent of the online search query, thus are inefficient for the online search query. 

% and are independent of the online search query.

% just as input data construction mentioned above is calculated offline, the quality assessment scores of all webpages are calculated offline too. The efficiency of the online ranking system is guaranteed. 

% To apply our layout-aware webpage quality assessment model in our online retrieval system, the new quality scores need to be loaded into the retrieval feature list.  

%% file: sections/6_evaluation/offline_evaluation.tex
\section{Offline Evaluation}\label{sec:offline_evaluation}
In this section, we conduct an offline evaluation of the proposed layout-aware webpage quality assessment model on the manually-labeled dataset from the search engine serves through offline experiments.

\begin{table*}[t]
\centering
\caption{Offline experimental results of different models.}
\resizebox{1\textwidth}{!}{
\begin{tabular}{l|c|c|ccc|ccc}
\toprule
\multirow{2}{*}{\bf Model} & \multirow{2}{*}{$\Delta$ \bf PNR} & \multirow{2}{*}{$\Delta$ \bf AUC (\%)} & \multicolumn{3}{c|}{\bf label 0} & \multicolumn{3}{c}{\bf label 1} \\ 
% \cline{4-9} 
& & & \multicolumn{1}{c}{$\Delta$ \bf P (\%)} & \multicolumn{1}{c}{$\Delta$ \bf R (\%)}  & $\Delta$ \bf F1 (\%)  & \multicolumn{1}{c}{$\Delta$ \bf P (\%)} & \multicolumn{1}{c}{$\Delta$ \bf R (\%)} &$\Delta$  \bf F1 (\%) \\ 
\midrule
TreeLSTM & 1.4 $\pm$ 0.01 & 14.83 $\pm$ 0.07 & \multicolumn{1}{c|}{6.25 $\pm$ 0.01} & \multicolumn{1}{c|}{\textbf{17.87 $\pm$ 0.07}} & 12.37 $\pm$ 0.03          & \multicolumn{1}{c|}{17.55 $\pm$ 0.06}          & \multicolumn{1}{c|}{-4.38 $\pm$ 0.06}          & 8.26 $\pm$ 0.05          \\ 

GIN & 2.76 $\pm$ 0.05  & 21.16 $\pm$ 0.20 & \multicolumn{1}{c|}{10.38 $\pm$ 0.39} & \multicolumn{1}{c|}{17.05 $\pm$ 1.17} & 14.01 $\pm$ 0.44 & \multicolumn{1}{c|}{21.23 $\pm$ 0.96} & \multicolumn{1}{c|}{7.07 $\pm$ 1.45} & 15.74 $\pm$ 0.26\\

GAT       & 2.92 $\pm$ 0.06   & 21.84 $\pm$ 0.23  & \multicolumn{1}{c|}{11.43 $\pm$ 1.03} & \multicolumn{1}{c|}{15.94  $\pm$ 2.21} & 13.91 $\pm$ 0.69 & \multicolumn{1}{c|}{20.91 $\pm$ 1.52}  & \multicolumn{1}{c|}{10.08 $\pm$ 3.40} & 16.93 $\pm$ 0.75 \\
\midrule
\midrule
GIN-NC &  2.64 $\pm$ 0.07    & 20.70 $\pm$ 0.30          & \multicolumn{1}{c|}{9.92 $\pm$ 1.42}          & \multicolumn{1}{c|}{17.30 $\pm$ 3.57}          & 13.86 $\pm$ 1.19          & \multicolumn{1}{c|}{21.15 $\pm$ 2.62}          & \multicolumn{1}{c|}{5.63 $\pm$ 5.15}          & 14.90 $\pm$ 1.17          \\
GAT-NC  & 2.75 $\pm$ 0.04                     & 21.17 $\pm$ 0.15          & \multicolumn{1}{c|}{10.41 $\pm$ 0.23}          & \multicolumn{1}{c|}{17.12 $\pm$ 0.75}          & 14.07 $\pm$ 0.31          & \multicolumn{1}{c|}{21.23 $\pm$ 0.67}          & \multicolumn{1}{c|}{7.13 $\pm$ 0.87}          & 15.83 $\pm$ 0.25          \\
\begin{tabular}[c]{@{}l@{}}Virt-GIN-NC\end{tabular} &  2.97 $\pm$ 0.04     & 21.95 $\pm$ 0.14          & \multicolumn{1}{c|}{11.26 $\pm$ 0.53}          & \multicolumn{1}{c|}{15.50 $\pm$ 1.27}          & 13.62 $\pm$ 0.44          & \multicolumn{1}{c|}{20.26 $\pm$ 0.86}          & \multicolumn{1}{c|}{9.88 $\pm$ 1.77}          & 16.53 $\pm$ 0.33          \\
\begin{tabular}[c]{@{}l@{}}Virt-GAT-NC\end{tabular} & 3.52 $\pm$ 0.03 & 23.56 $\pm$ 0.08          & \multicolumn{1}{c|}{12.55 $\pm$ 0.48}          & \multicolumn{1}{c|}{17.41 $\pm$ 1.23}          & \textbf{15.23 $\pm$ 0.43} & \multicolumn{1}{c|}{\textbf{23.38 $\pm$ 1.04}} & \multicolumn{1}{c|}{12.29 $\pm$ 1.60}          & 19.34 $\pm$ 0.25          \\
\midrule
\midrule
Virt-GIN                                                         & 3.11 $\pm$ 0.03                 & 22.37 $\pm$ 0.10          & \multicolumn{1}{c|}{11.79 $\pm$ 0.41}          & \multicolumn{1}{c|}{14.96 $\pm$ 1.49}          & 13.57 $\pm$ 0.61          & \multicolumn{1}{c|}{20.17 $\pm$ 1.20}          & \multicolumn{1}{c|}{11.38 $\pm$ 1.56}          & 17.12 $\pm$ 0.14          \\
%  |  |  |  |  |  |  | 
Virt-GAT  & \textbf{3.71 $\pm$ 0.10}        & \textbf{24.08 $\pm$ 0.24} & \multicolumn{1}{c|}{\textbf{13.37 $\pm$ 0.57}} & \multicolumn{1}{c|}{16.18 $\pm$ 1.13}          & 14.96 $\pm$ 0.35          & \multicolumn{1}{c|}{22.66 $\pm$ 0.79}          & \multicolumn{1}{c|}{\textbf{14.67 $\pm$ 1.71}} & \textbf{19.97 $\pm$ 0.29} \\
\bottomrule
\end{tabular}}                                   
\label{table:offline_results}
\end{table*}

\subsection{Dataset}
% longer
To evaluate the proposed method, 
% we evaluate the model on the dataset collected from the real production environment
we first collect a set of webpages from our database, which stores the real webpages that our search engine serves.
Next, we manually label all the collected webpages
% add, 'time, condition' need review
% We collect the dataset from [Jan. 2021 to Jun. 2021] provided by Baidu search engine, and only keep those samples with more than xxx. 
% Since only xxx over a period of six months is likely to be an occasional and unintentional click, which could reduce the quality of the dataset.
% The dataset is labeled
on our crowdsourcing platform, where a group of experts are required to assign low-quality (0) or high-quality (1) to each of the given webpage. 
% an integer score that varies from 0 to 1 to each webpage without considering the query of search engine.
% The score represents whether the layout of the webpage w.r.t. the query is 
% We denote this dataset as \textbf{Manual} dataset.
In our experiments, we use 600,000 webpages for training and 20,000 webpages for testing.
% The score represents whether the layout of the webpage w.r.t. the query is chaos (0), common (0), clearly structured and informative (2) or elegant (3). 

\subsection{Evaluation Metrics}

\noindent\textbf{Positive-Negative Ratio (PNR).} We use PNR to measure the consistency between manual quality labels and the scores estimated by the model.
% We report the positive-negative ratio
% (PNR) on Manual dataset. 
% Positive-negative ratio
% (PNR) 
In particular, by enumerating all the pairs of webpages in the dataset (i.e., $D$), PNR
can be formally defined as
\begin{equation}
PNR=\frac{\sum_{d_{i}, d_{j} \in D} \mathbb{I}\left(y_{i}>y_{j}\right) \cdot \mathbb{I}\left(f\left(d_{i}\right)>f\left(d_{j}\right)\right)}{\sum_{d_{i^{\prime}}, d_{j^{\prime}} \in D} \mathbb{I}\left(y_{i^{\prime}}>y_{j^{\prime}}\right) \cdot \mathbb{I}\left(f\left(d_{i^{\prime}}\right)<f\left(d_{j^{\prime}}\right)\right)},
\end{equation}
where $\mathbb{I}$ is an indicator function, i.e., $\mathbb{I}\left(a>b\right)=1$, if $a>b$, and $0$ otherwise. Here, $f(d_i)$ represents the quality score of a webpage $d_i$ estimated by the model. Higher PNR value indicates better performance of the model.
% PNR measures the consistency between the manual labels and the model scores. 
% We report the PNR values under the same query for all webpages in the experiments.

\noindent\textbf{Area Under Curve, Precision, Recall, F1-Score.} We also report 
% the commonly metrics used to measure the performance of a classification model, i.e., 
Area Under Curve (AUC), Precision (P), Recall (R) and F1-Score (F1) to evaluate our proposed model. Precision and recall are often in tension, that is, improving precision typically reduces recall and vice versa. F1-Score combines them to one performance metric. Area under curve summarizes the trade-off between the true positive rate and false positive rate for a predictive model using different probability thresholds.

% revise baseline part
\subsection{Compared Baselines and Our Approach}
To validate the effectiveness of our layout-aware webpage quality model, we conduct experiments on several related 
% We compare our proposed method with several
\textbf{baseline models:} \textbf{TreeLSTM} \cite{tai2015improved}, a standard LSTM architecture designed for tree-structured network topologies. \textbf{GIN} \cite{xu2018powerful} introduces a learnable parameter to adjust the weight of the central node. \textbf{GAT} \cite{velivckovic2017graph} leverages the attention mechanism to improve neighbor aggregation scheme. 
\textbf{Our proposed models:} \textbf{Virt-GIN} has a more expressive readout mechanism by adding the virtual node $\vec{h}_{v}$ to GIN model. \textbf{Virt-GAT} is our approach similar to virt-GIN model, i.e., a GAT model with virtual node. \textbf{Models-NC:} Note that all the above-mentioned models use category information as proposed in Section \ref{sec:model}. To further clarify the influence of category in the model, we also include four variants without using category information, which is denoted with a suffix Non-Category (\textbf{-NC}).

% In addition, we also compare our proposed method with online baseline, 
We compare our proposed method with the online method in our search system, which is the quality assessment model that was previously served online in our search engine. This can clearly illustrate the improvement brought by the proposed solution for our search engine.

\subsection{Experimental Settings}
% The dataset train/validation/test split is 0.8/0.1/0.1 for all the compared methods. 
In our experiments, Adam is selected as the optimizer. We use the following hyper-parameters: embedding size (64), number layers (5), dropout probability (0.2), batch size (32), learning rate (0.0001) for GNN models, and train epochs (25).
% We train each methods for 25 , and then the test ROC-AUC score is chosen at the best validation epoch. 
As for the TreeLSTM model, we set the embedding size (64), dropout probability (0.5), batch size (128), learning rate (0.0001), epochs (25) for it. We run 5 experiments with different random seeds for all models mentioned above. The final result we reported is the mean test AUC, Precision, Recall, F1-Score and their corresponding standard deviation.
All the above mentioned GNN models are implemented by Paddle Graph Learning (PGL)\footnote{\url{https://github.com/PaddlePaddle/PGL}}, an efficient and flexible graph learning framework.

\begin{table*}[t]
\centering
\caption{The influence of layer number on virt-GAT.}
{
\begin{tabular}{c|c|ccc|ccc}
\hline
\toprule
\multirow{2}{*}{\bf \#Layers} & \multirow{2}{*}{$\Delta$ \bf AUC (\%)} & \multicolumn{3}{c|}{\bf label 0} & \multicolumn{3}{c}{\bf label 1} \\  
 &                      & \multicolumn{1}{c}{$\Delta$ \bf P (\%)}                & \multicolumn{1}{c}{$\Delta$ \bf R (\%)}                &$\Delta$  \bf F1 (\%)               & \multicolumn{1}{c}{$\Delta$ \bf P (\%)}                & \multicolumn{1}{c}{$\Delta$ \bf R (\%)}                & $\Delta$ \bf F1 (\%)               \\ 
                          \midrule
\bf 1 & -3.41 $\pm$ 0.23     & \multicolumn{1}{c|}{-3.43 $\pm$ 0.92} & \multicolumn{1}{c|}{0.29 $\pm$ 2.49} & -1.48 $\pm$ 0.85 & \multicolumn{1}{c|}{-2.49 $\pm$ 1.82} & \multicolumn{1}{c|}{-8.52 $\pm$ 3.42} & -5.24 $\pm$ 0.66 \\
\midrule
\bf 3 & -0.38 $\pm$ 0.27     & \multicolumn{1}{c|}{-0.56 $\pm$ 0.64} & \multicolumn{1}{c|}{-0.37 $\pm$ 2.06} & -0.46 $\pm$ 0.86 & \multicolumn{1}{c|}{-0.77 $\pm$ 1.79} & \multicolumn{1}{c|}{-1.19 $\pm$ 2.16} & -0.98 $\pm$ 0.45 \\
% \midrule
% \bf 5  & 84.18 $\pm$ 0.24     & \multicolumn{1}{c|}{86.81 $\pm$ 0.57} & \multicolumn{1}{c|}{80.17 $\pm$ 1.13} & 83.35 $\pm$ 0.35 & \multicolumn{1}{c|}{62.75 $\pm$ 0.79} & \multicolumn{1}{c|}{73.24 $\pm$ 1.71} & 67.57 $\pm$ 0.29 \\
\midrule
\bf 7 & 0.07 $\pm$ 0.22    & \multicolumn{1}{c|}{0.10 $\pm$ 0.80} & \multicolumn{1}{c|}{0.36 $\pm$ 1.77} & 0.23 $\pm$ 0.61 & \multicolumn{1}{c|}{0.48 $\pm$ 1.41} & \multicolumn{1}{c|}{0.08 $\pm$ 2.43} & 0.30 $\pm$ 0.42 \\ 
\bottomrule
\end{tabular}}                                       
\label{table:layers_offline_results}
\end{table*}

\subsection{Offline Experimental Results}
We report the offline experimental results of the proposed model and all baseline models. And all results are the absolute improvement value over the online method, i.e., the method that is used in the system before deploying the layout-aware webpage quality assessment model. Details are shown in Table \ref{table:offline_results}, from where we have the following \textbf{key findings}:
\begin{itemize}
\item We can clearly see that our layout-aware webpage quality model can beat the online method by large margins on all metrics e.g., $\Delta AUC=24.08\%$, $\Delta F1=14.96\%$ (label0) and $\Delta F1=19.97\%$ (label1). Especially for PNR, where the improvement is 3.71. These tell us that the proposed model prefers high-quality results.
\item By applying the proposed readout function, the model can have a significant improvement on all metrics. Especially, the new readout mechanism is able to improve PNR by a margin of 0.38 and 0.96 based on GIN and GAT, respectively. Moreover, we also observe that the improvement of both virt-GIN and virt-GAT over GIN and GAT is considerable for high-quality webpage (label1), in terms of recall ($\Delta_{(Virt\_GAT, GAT)}=4.59\%$, $\Delta_{(Virt\_GIN, GIN)}=2.22\%$). All these phenomena show that our readout mechanism is capable of improving the model's performance.
\item Comparing the results of the two models whether apply the category-aware optimization strategy (w,r,t., GIN-NC vs. GIN, Virt-GIN-NC vs. Virt-GIN, GAT-NC vs. GAT, Virt-GAT-NC vs. Virt-GAT), we can come to the conclusion that all methods with the proposed category-aware optimization have better performance than their backbone models, in terms of PNR and AUC. Although a few models obtain lower values on a few metrics (e.g., the relative F1-score of Virt-GAT-NC on label0 is 15.23\% while Virt-GAT is 14.96\%, the relative precision of Virt-GAT-NC is 23.38\% but Virt-GAT is 22.66\%), the models with category-aware optimization show more robust performance considering all metrics.
\item The performance on different GNN models is better than TreeLSTM, model Virt\_GAT is the most significant, Compare with Virt\_GAT and TreeLSTM, $\Delta PNR=2.31$, $\Delta AUC=9.25\%$. For high-quality webpage (label1) $\Delta R=14.67\%$. These large margins suggest that our model is more expressive than TreeLSTM, although TreeLSTM is specifically designed for tree-structured network topologies.
\end{itemize}

Overall, our proposed model is able to gain superior performance on webpage assessment tasks through the improved readout mechanism and category-aware optimization and can beat the online baseline by a significant margin.

\subsection{Varying the number of GNN layer}
In general, a webpage is represented as a DOM tree. Its depth determines how many layers of GNN are needed to obtain information from the root node to the leaf nodes. 
However, as the number of GNN layers increases, the computational complexity will increase.
Therefore, we provide an experiment on our model virt-GAT to verify the influence of the number of layers on the experimental results, as shown in Table \ref{table:layers_offline_results}. All results are reported as absolute improvement values over 5-layer virt-GAT model.
As seen from the table, the more layers, the higher the AUC score can be reached. When the number of layers comes to 7, the improvement is slight.
% However, compared with the 5-layer virt-GAT model, the improvement of 7-layer virt-GAT model is not significant. 
As it is important to trade off the efficiency and effectiveness for large search system, we use 5-layer GNN models on online evaluation which can maintain the experimental effect while reducing the amount of calculation.

%% file: sections/6_evaluation/online_evalution.tex
\section{Online Evaluation}\label{sec:online_evaluation}
To investigate the impact of our proposed quality assessment model to the search engine, we deploy the new model and conduct online experiments to compare it with the old retrieval system. Specifically, we conduct a manual evaluation on the final ranking results with some real user-generated queries. This directly reflects the quality of the results exposed to the end users.
% \subsection{Data Preparation}

We log a set of (million-scale) online queries and the corresponding final impressions, i.e., the top-ranked web documents in the final ranking stage, by individually using the layout-aware webpage quality assessment model and the old retrieval systems. Note that the data logging is conducted by multiple rounds to eliminate randomness. We filter out examples in which queries have identical impressions between the two systems, and then utilize the rest for the manual evaluation. Note that, considering the extremely high cost of the manual evaluation, we randomly generate thousands of data and eventually send it to experts for evaluation, so as to control costs while validating the effectiveness of the proposed model.

%  The left component shows three webpage and their quality scores, while the right component reports position changes of two webpages between new and old retrieval system.

\subsection{Online Experimental Metrics}
As mentioned in Section \ref{sec::deployment}, our proposed quality assessment model works in Baidu retrieval system. The online experiments major focus on the end-to-end evaluation, the metrics are often used to measure the effectiveness of information retrieval system. Details are as follows:
\paragraph{\textbf{Discounted Cumulative Gain (DCG)}}
We first log a dataset and manually label the data with 0 to 4 grades, and then report the relative improvement w.r.t. the average DCG over the top-4 final results of all queries. The formula of DCG accumulated at a particular rank position $\mathrm{p}$ is defined as
\begin{equation}
\label{eq:DCG}
    \mathrm{DCG}_{\mathrm{p}}=\sum_{i=1}^{p} \frac{2^{rel_{i}}-1}{\log _{2}(i+1)},
\end{equation}
where $rel_{i}$ indicates the manually label of $i$-th webpage. 

Additionally, we also report the relative improvement of DCG for the low quality ranking result w.r.t., manually label is 0/1. 
\paragraph{\textbf{Side-by-side Comparison}}
Besides, we also conduct a side-by-side comparison between the two systems. We log another dataset and require the human experts to judge whether the new system or the base system gives better results that satisfy intentions of users. 
Here, the relative gain is measured Good vs. Same vs. Bad (GSB) as
\begin{equation}
\label{eq:gsb}
    \Delta_{GSB}=\frac{\text{\#Good}-\text{\#Bad}}{\text{\#Good}+\text{\#Same}+\text{\#Bad}},
\end{equation}
where $\text{\#Good}$ (or \text{\#Bad}) indicates the number of queries that the new system provides better (or worse) final results.
\begin{table}[t!]
\centering
\caption{Discounted cumulative gain on manual evaluation.}
% \resizebox{0.48\textwidth}{!}
{
\begin{tabular}{lccc}
\toprule & Rand-Query & Tail-Query & Same-Quality\\
\midrule
$\Delta_{DCG}$ & +0.19\% & +0.42\% & -\\
\midrule
DCG\_0/1 ratio & -0.63\% & -0.56\% & -\\
\bottomrule
\end{tabular}}
\label{table:dcg_results}
\end{table}

Node that we not only measure the final results but also measure the webpage quality when the relative result of two webpage is Same.

\begin{figure*}[t!]
	\centering
	\subfigure[Offline quality assessment]{\includegraphics[width=0.642\linewidth]{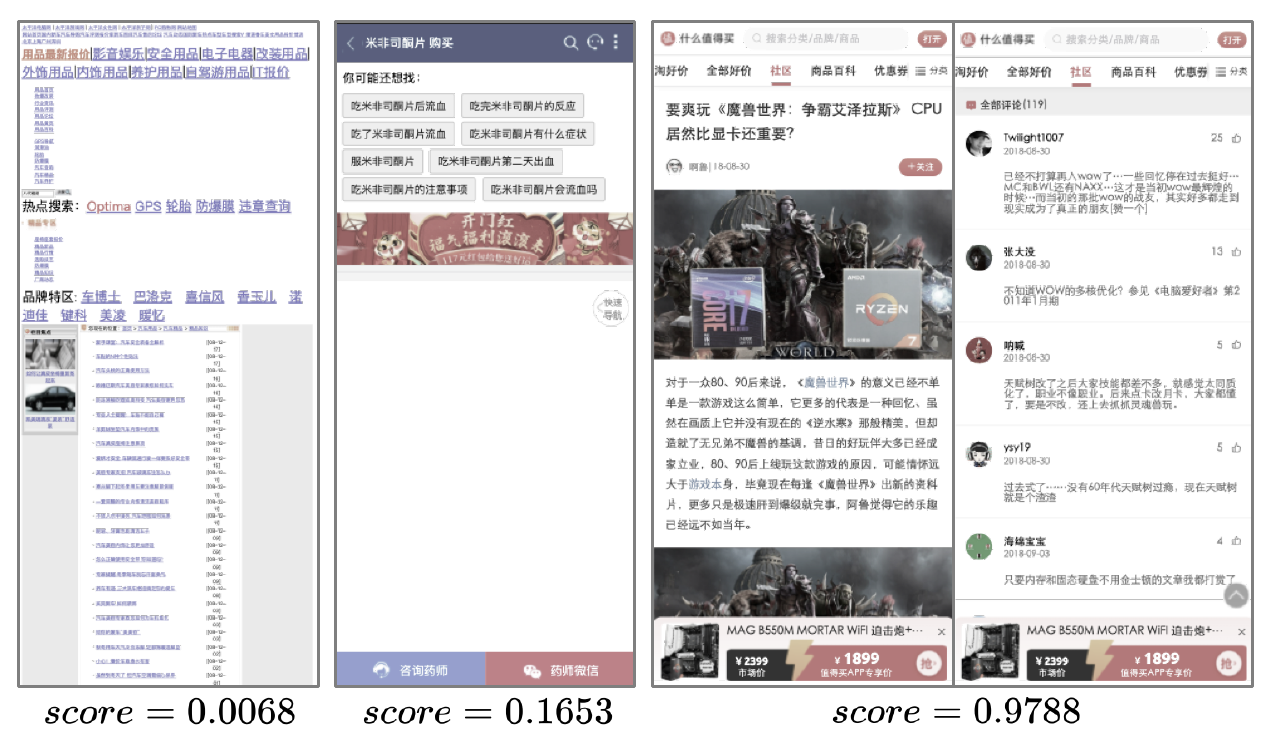}\label{fig:subfigure_offline_case}}
    \subfigure[Online position changes]{\includegraphics[width=0.327\linewidth]{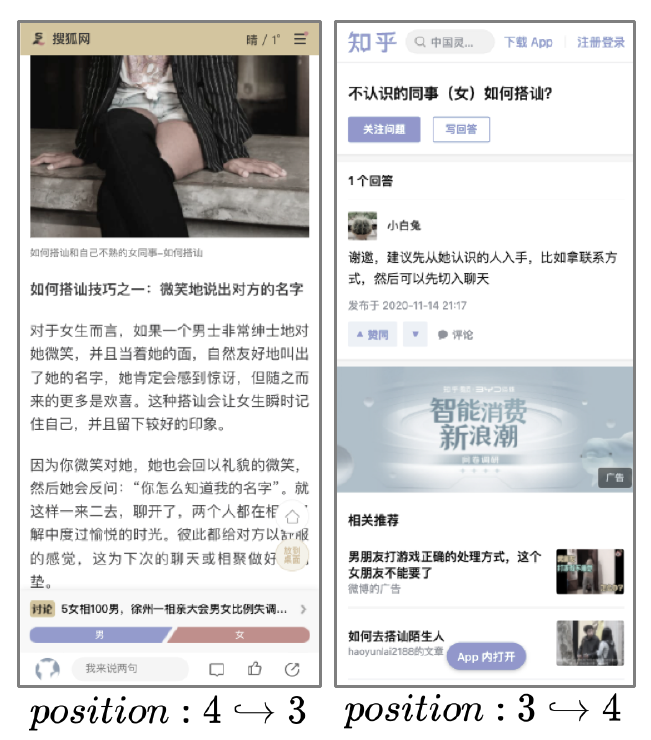}\label{fig:subfigure_online_case}}
	\caption{The overview of case study.}
	\label{fig:case_study}
\end{figure*}

\subsection{Online Experimental Results}
% \ref{table:online_results}
The relative improvement validated by manual evaluation is given in Table~\ref{table:dcg_results} and \ref{table:gbs_results}, where we can summarize observations as below:
\begin{itemize}
    \item By applying our quality assessment model, the system can significantly outperform the base system. Especially for DCG\_0/1 ratio, the relative improvement values are respectively $-0.63\%$, $-0.56\%$ for rand query and tail query. This shows that our proposed method can better filtrate retrieval results with low DCG scores, which is very helpful in improving the user experience for real-world search engine.
    \item The conventional case-by-case comparison also has significant improvement over the base system, especially for the rand query ($\Delta_{GSB}=+4.1\%$). This tells us that user experience can be improved by taking into account the web page quality in search system.
    \item In addition, we can observe that with comparable relevance, the GSB value of the quality improvement is $\Delta_{GSB}=+5.13\%$. This intuitively shows that our new system can provide higher quality search results based on the guaranteed relevance of search results.
\end{itemize}

\begin{table}[t!]
\centering
\caption{Side-by-side comparison on manual evaluation.}
% \resizebox{0.48\textwidth}{!}
{
\begin{tabular}{lccc}
\toprule & Rand-Query & Tail-Query & Same-Quality\\
\midrule
$\Delta_{GSB}$ & +4.10\% & +0.52\% & +5.13\%\\
\bottomrule
\end{tabular}}
\label{table:gbs_results}
\end{table}

Moreover, we perform the statistical test to estimate whether the experimental results is statistically significant. The p-value of DCG rand and tail query are $0.0613$ and $0.1276$, respectively. The p-value approximates the significance level that is set in our retrieval system, which can demonstrate that our experimental results are statistically significant.

Overall, the online experimental results show that our proposed layout-aware quality assessment model can effectively improve the performance of real-world ranking system.

%% file: sections/7_case_study.tex
\section{Case Study}
In this section, we present an illustration that includes the offline quality assessment score of webpage and online position changes of web pages. These end-to-end cases are shown in Figure~\ref{fig:case_study}.

\subsection{Offline Quality Assessment}
% \noindent\textbf{Offline quality assessment.}
% As shown in Section~\ref{sec:offline_evaluation}, our layout-aware quality assessment model can gain significant performance on the manually dataset. 
In Figure~\ref{fig:subfigure_offline_case}, we present three webpages with different layout styles and their quality assessment scores. 

\textbf{The first webpage} has a chaotic layout, elements in this webpage are unreasonable. It affects the user's normal browsing and is very difficult for user to obtain information from this webpage. Our quality assessment model marks this webpage as low quality ($score=0.0068$). This extremely low score will be considered by the ranking system to lower its ranking position.

\textbf{The second webpage} also has low quality, different from the chaotic layout of the first webpage, it has a normal layout. However, considering that it contains a tiny amount of information (almost no valuable information), it should be presented to the user with a very small probability. The ranking system can judge this by our quality assessment model score of $0.1653$.

Unlike the previous two webpages, \textbf{the third one} is high-quality. It is carefully laid out and informative, and quality score is $0.9788$, which will help the ranking system raise its ranking position.

\subsection{Online Position Changes}
% \noindent\textbf{Online Position Changes}.
The case shown in Figure~\ref{fig:subfigure_online_case} comes from Section~\ref{sec:online_evaluation}. Under the same query, these two webpages swapped positions in the new and old systems, The position of the left webpage in new system is $3$-th but $4$-th in the old system. Comparing the two webpages, we can observe that the left webpage (quality score is $0.5623$) contains a rich amount of information but the right one (quality score is $0.2415$) does not. This phenomenon demonstrates that online ranking system has adopted our model's recommendations to provide users with higher quality webpage, which can greatly improve the user experience.

%% file: sections/8_conclusion.tex
\section{Conclusion and Future Work}
In this paper, we propose a layout-aware webpage assessment model to suggest ranking system providing webpages with higher quality. We not only enhance GAT with the read mechanism but also carefully design the features for improving the quality assessment on the webpages. In addition, taking into account the particularity of real-world data, we utilize the category of webpage for optimization. Both input data construction and model calculation are offline, which guarantees the efficiency of the ranking system. We developed and deployed the model in Baidu Search.
% , which is highly effective in conducting high-quality ranking for web search. 
Extensive offline and online experiments have shown that the ranking system can significantly improve the effectiveness and general usability of the search engine. In future work, we will explore the heterogeneous GNN architecture to model the multiple graph-based information of webpages. It is interesting to improve the construction method of layout and enhance the representation of nodes/edges with self-supervised contrastive pre-training techniques.